\documentstyle[12pt,epsf]{article}
\newcommand{\agt}{\mathrel{\raisebox{-.6ex}{$\stackrel{\textstyle>}{\sim}$}}}
\def\overlay#1#2{\ifmmode \setbox 0=\hbox {$#1$}\setbox 1=\hbox to\wd 0{\hss
$#2$\hss }\else \setbox 0=\hbox {#1}\setbox 1=\hbox to\wd 0{\hss #2\hss }\fi
#1\hskip -\wd 0\box 1}
\begin{document}

\font\fortssbx=cmssbx10 scaled \magstep2
\hbox to \hsize{
\hbox{\fortssbx University of Wisconsin - Madison}
\hfill$\vcenter{\hbox{\bf MADPH-95-920}
                \hbox{\bf RAL-TR-95-075}
                \hbox{\bf IUHET-324}
                \hbox{\bf hep-ph/9512325}
                \hbox{December 1995}}$ }

\vspace{.75in}

\begin{center}{\large\bf
Thresholds in $\alpha_s$ evolution and the $p_T$ dependence of jets}\\[.6in]
V. Barger$^{(a)}$, M.S. Berger$^{(b)}$, and R.J.N. Phillips$^{(c)}$\\[.3in]
\it $^{(a)}$Physics Department, University of Wisconsin, Madison, WI 53706,
USA\\
$^{(b)}$Physics Department, Indiana University, Bloomington, IN 47405, USA\\
$^{(c)}$Rutherford Appleton Laboratory, Chilton, Didcot, Oxon OX11 0QX, UK
\end{center}

\vspace{.75in}

\begin{abstract}
  We point out that high-mass thresholds in the evolution of the
strong-interaction coupling parameter $\alpha_s$, due to
gluinos, squarks and possible new heavy quarks, could introduce
appreciable corrections to the transverse momentum dependence of
jet production at the Tevatron.  If the new thresholds were near
scale $\mu = 200$~GeV, then within the limits of asymptotic
freedom they could introduce up to $11\%$ increase in
$\alpha_s(\mu)$ (and hence $23\%$ increase in jet production)
at scale $\mu=500$ GeV, compared to Standard Model extrapolations.
\end{abstract}

\thispagestyle{empty}
\newpage

The CDF experiment at the Fermilab Tevatron collider has recently
reported that inclusive jet production in $p\bar p$ collisions
appears to exceed next-to-leading order QCD expectations in the
transverse momentum range $200 < p_T(j) < 420$ GeV \cite{cdf}.
It has been suggested that the discrepancy might be explained
by a modified gluon distribution in the nucleon \cite{cteq}.
However, it is also possible that some new physics may be playing
a significant role.  In the present note we point out that
high-mass thresholds due to gluinos $({\tilde g})$
and squarks $({\tilde q})$, plus possible new
non-standard quarks $(Q)$ and squarks $({\tilde Q})$,
would change the evolution of the
QCD coupling parameter $\alpha_s(\mu)$ at large values of the scale
$\mu \agt m_{\tilde g}, m_{\tilde q}, m_Q, m_{\tilde Q}$.
This effect could give a significant enhancement above standard
expectations, both for $\alpha_s(\mu)$ at large $\mu$ and hence
also for the jet production cross section at large $p_T$.

The evolution equation for $\alpha_s(\mu)$ at one loop can
be written
\begin{equation}
{d \over dt}[\alpha_s(\mu)]^{-1} = {b_3 \over 2\pi},
\end{equation}
with
\begin{equation}
b_3 = 11 - {2 \over 3}n_f,
\label{eq:b3sm}
\end{equation}
where $t=\ln \mu$ and $n_f$ is the number of quark flavors that are active,
i.e. that have $\mu > m_q$  \cite{marc}.
In a supersymmetric (SUSY) model, $b_3$ reduces by two units above
the gluino threshold and by ${1 \over 3}$ above each successive squark
threshold (assuming degenerate L- and R-chiral squarks), giving
\begin{equation}
b_3 = 9 - {2 \over 3}n_f - {1 \over 3}\tilde n_f,
\label{eq:b3gen}
\end{equation}
where we assume the gluino is active and $\tilde n_f$ denotes the
number of active squark flavors.
It is clear that as the scale $\mu$ increases,
successively more thresholds are crossed and the rate of evolution
of $[\alpha_s(\mu)]^{-1}$ decreases.  Accordingly, the rate of
fall of $\alpha_s(\mu)$ as $\mu$ increases (asymptotic freedom)
is reduced with each successive threshold.
Further new physics, for example new iso-singlet quarks $Q_i$ and
their SUSY partners $\tilde Q_i$ (that appear naturally in the [27]-plet
of $E_6$ symmetry \cite{rosner,desh}) would reduce $b_3$ even more
\cite{bbp}; each counts the same as a conventional
quark or squark in Eq.(\ref{eq:b3gen}).  Alternatively, a fourth
generation would add two quark plus two squark flavors\cite{gunion}.
With three new quarks plus squarks the value
$b_3=0$ is reached and we can go no further without breaking
asymptotic freedom (AF).  Although AF is not an inviolable principle,
it is an attractive feature; we therefore limit our present
illustrations to cases up to the AF limit.

The solid curve in Fig.1 shows the one-loop evolution of $\alpha_s(\mu)$
versus $\mu$ for the standard model (SM) colored-particle content.
The dotted curve shows the change to the minimum supersymmetric
extension (MSSM), assuming for simplicity a common squark/gluino
threshold at $\mu = m_{\tilde q} = m_{\tilde g} = 200$ GeV.
The dot-dashed curve shows the effect of adding three singlet quarks
with the same threshold; the case of one or two singlet quarks can
be obtained by interpolation between dotted and dot-dashed curves.
Finally the horizontal dashed curve shows that adding three singlet
squarks too, with the same threshold, brings us to the AF limit.
The dashed curve lies above the solid SM curve by a factor 1.11
at scale $\mu=500$ GeV.

The inclusive jet production cross section $\sigma(p\bar p\to jX)$
at the Tevatron is based on $2\to 2$ QCD parton subprocesses
that are proportional to $[\alpha_s(\mu)]^2$.  A common choice
for the scale $\mu$ here is $\mu = p_T(j)/2$, in terms of the
jet transverse momentum $p_T$; in this case the new threshold
effects of Fig.1 would only be felt for $p_T > 400$ GeV, i.e.\
mostly beyond the range of the reported Tevatron data\cite{cdf}.
However, for the choice $\mu = p_T(j)$ instead, the effects of
Fig.1 would give an enhancement for $p_T>200$ GeV; at $p_T=500$ GeV
this enhancement ranges between $12\%$ (for the MSSM case) and
$23\%$ (for the AF limit case).  The relationship between
$\mu$ and $p_T$ could be determined in principle by
higher-order calculations of jet production, including the
physics beyond the Standard Model in the loops, but for the
present it remains an open question.

We conclude that new thresholds in the evolution of $\alpha_s$
are capable of providing substantial increases in the jet
hadroproduction cross section at large $p_T(j)$, and may be
significant contributors to the reported effect.

\section*{Acknowledgments}

This research was supported in part by
the U.S.~Department of Energy under Grant Nos.~DE-FG02-95ER40896 and
DE-FG02-91ER40661 and in part by the University of Wisconsin
Research Committee with funds granted by the Wisconsin Alumni Research
Foundation.

\section*{Figure captions}

\noindent
Fig.1.  The one-loop evolution of $\alpha_s(\mu)$ versus $\mu$
is compared in various scenarios, normalized to $\alpha_s(M_Z)=0.12$.
The solid curve is the SM prediction, with $m_t=175$ GeV.  The dotted
curve is the MSSM prediction, with a common squark/gluino threshold
at $\mu=200$ GeV. The dot-dashed curve shows the effect of three
singlet quarks with common mass $m_Q=200$ GeV.  The dashed curve,
which reached the AF limit, shows the further effect of three singlet
squarks if $m_{\tilde Q}=200$ GeV.

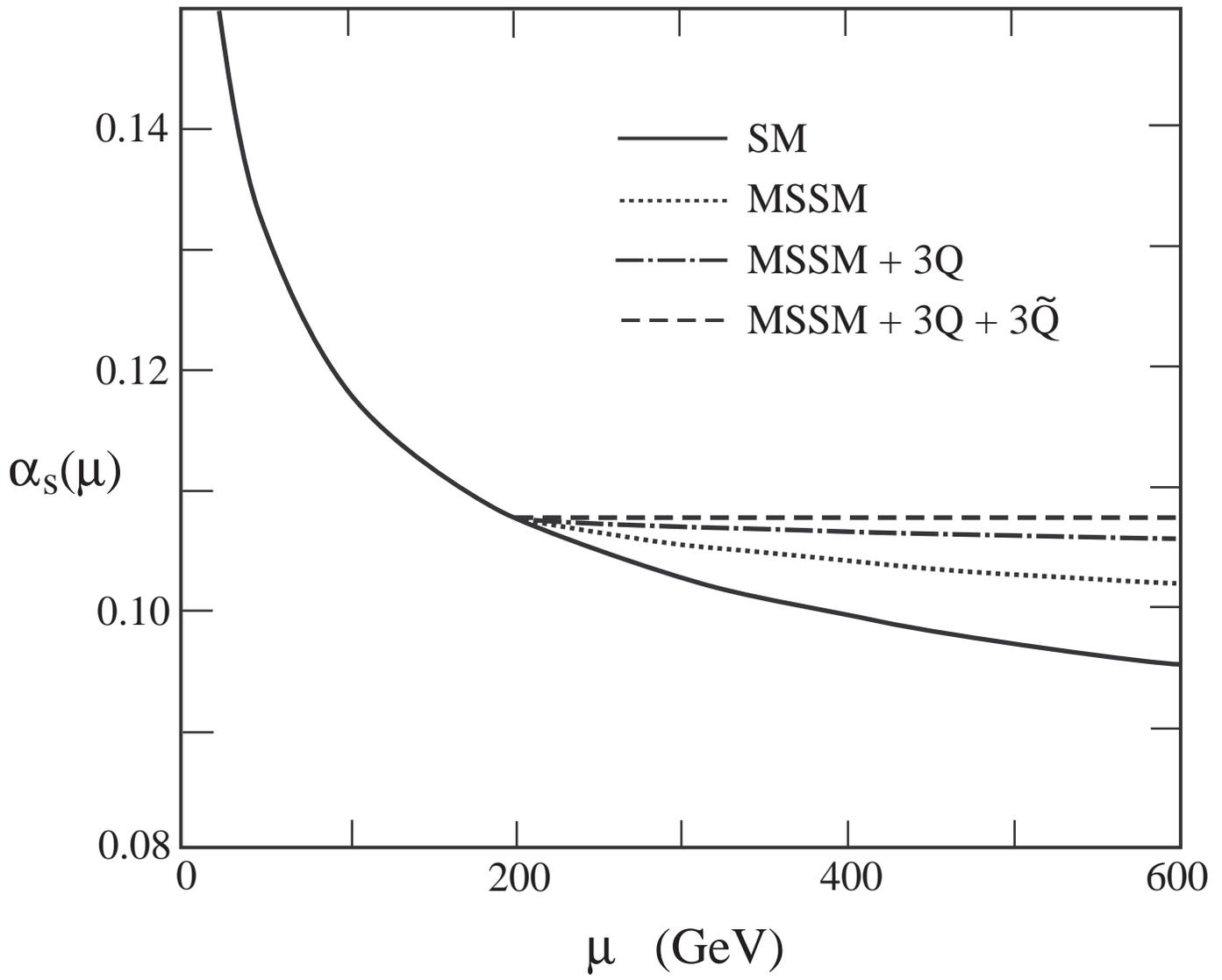
\end{document}